\documentclass[%
 reprint,
 amsmath,amssymb,
 aps,
]{revtex4-2}

\usepackage[utf8]{inputenc}  
\usepackage[T1]{fontenc}     

\usepackage{physics}
\usepackage{graphicx}
\usepackage{dcolumn}
\usepackage{bm}
\usepackage{booktabs}
\usepackage{makecell}
\usepackage{siunitx}  
\usepackage[hidelinks]{hyperref}
\begin{document}

\title{Systematic optimization of resonance parameters in a transport approach}

\author{Carl B. Rosenkvist$^{1,2,3}$ Hannah Elfner$^{3,1,2,4}$}

\affiliation{$^1$Frankfurt Institute for Advanced Studies, Ruth-Moufang-Strasse 1, 60438 Frankfurt am Main, Germany}
\affiliation{$^2$Institute for Theoretical Physics, Goethe University, Max-von-Laue-Strasse 1, 60438 Frankfurt am Main, Germany}
\affiliation{$^3$GSI Helmholtzzentrum für Schwerionenforschung, Planckstr. 1, 64291 Darmstadt, Germany}
\affiliation{$^4$Helmholtz Research Academy Hesse for FAIR (HFHF), GSI Helmholtz Center, Campus Frankfurt, Max-von-Laue-Straße 12, 60438 Frankfurt am Main, Germany}

\date{\today}

\begin{abstract}
This study optimizes resonance parameters responsible for strangeness production in the SMASH (Simulating Many Accelerated Strongly-interacting Hadrons) transport model using a genetic algorithm. By fitting resonance parameters to experimental data on exclusive strangeness cross-sections at low energies, we significantly improve the model's accuracy, especially in pion-proton interactions. Our approach explores how machine learning tools can be used for precise resonance tuning in transport approaches. 

\end{abstract}

\maketitle


\section{Introduction}

One of the primary goals of heavy-ion collisions is to explore the quantum chromodynamics (QCD) phase diagram, which describes the behavior of strongly interacting matter under extreme conditions of temperature and density. Both theoretical and experimental studies suggest the existence of a phase, known as the quark-gluon plasma (QGP), where quarks and gluons are nearly deconfined\cite{Harris_1996}. This state of matter is expected to emerge at high temperatures and/or densities, providing a unique environment to study the fundamental properties of QCD.

Strangeness plays a crucial role in this investigation because it is conserved under the strong interaction and must have been created during the collision. Its presence can offer significant insights into the properties of the quark-gluon plasma (QGP).

Various models have been developed to describe strange quark production under different energy and density conditions. At higher energies, event generators such as Pythia/Angantyr\cite{10.21468/SciPostPhysCodeb.8, Angantyr} utilize the Lund string fragmentation model to simulate hadronization, including the production of strange quarks. Recently, the string framework has been extended with rope hadronization \cite{Bierlich_2018}, a feature added to these models to address strangeness enhancement.

In contrast, models that focus on the thermodynamics of heavy-ion collisions, such as hydrodynamic and statistical hadronization models, describe strange quark production through notably gluon interactions\cite{PhysRevLett.48.1066}. These approaches focus on the collective behavior of the medium rather than on individual hadronization processes. 

At lower energies, where hadronic dynamics dominate, transport approaches provide a more suitable description of the observed phenomena. For example, the Parton-Hadron-String Dynamics (PHSD) approach, which models both nucleon resonances and color-dipoles (strings) \cite{Bratkovskaya:2019gms}, incorporates in-medium effects on strangeness production. 

This is achieved through modifications to the antikaon potential, spectral functions, and reaction cross sections, which depend on baryon density, temperature, and antikaon momentum in the nuclear medium \cite{Song_2021}. This approach has emphasized the significance of in-medium effects on strangeness in nuclear matter.

Complementing the medium effects, the vacuum properties of resonances, such as pole positions and branching ratios, also play a significant role in strangeness production. In transport models like Ultra-relativistic Quantum Molecular Dynamics (UrQMD), strangeness production at lower energies is included through resonance decay \cite{URQMD}. These resonances are modeled by their vacuum properties tuned to experimental data.

A newer transport model, Simulating Many Accelerated Strongly-interacting Hadrons (SMASH), shares similarities with UrQMD but differs in the selection of resonances and resonance parameters \cite{SMASH-3.1}. At higher energies, both SMASH and UrQMD employ string fragmentation handled by Pythia.

However, resonance properties, such as branching ratios, are often poorly constrained by experimental data, leading to uncertainties in model predictions. This is demonstrated in the UrQMD study \cite{pionpaper}, which investigated model stability under variations in resonance mass and widths. Their findings showed that at lower energies, where resonance dynamics dominate, transport models are sensitive to resonance parameters. In SMASH, resonance properties have been tuned to experimental data \cite{vinzent}, but a systematic approach to investigate different parameter sets has not yet been applied.

To address this challenge, we propose a more systematic approach by leveraging experimental data on exclusive strangeness cross-sections at low energies to fit the resonance parameters. In this paper, we utilize a genetic algorithm for parameter optimization and apply it within the SMASH transport model to simulate the cross-sections and refine the resonance properties.



The remainder of this paper details our simulation set-up, including the use of SMASH and the development of an emulation method. We then describe the application of genetic algorithms to optimize resonance parameters. This is followed by a presentation of our results, where we compare simulation outcomes with experimental data. We discuss the implications of our findings and suggest directions for future research. Finally, we conclude by summarizing the key contributions and significance of our work.

\section{Resonance production in SMASH}
SMASH is a relativistic hadron transport approach for heavy-ion collisions, incorporating resonances up to a mass of $\sim 2\,\text{GeV}$ as degrees of freedom at lower energies. At higher energies, strings are formed and then fragmented with \texttt{Pythia}. 

In SMASH, resonances are modeled by a spectral function $\mathcal{A}$, represented as a relativistic Breit-Wigner distribution.

\begin{equation}
\mathcal{A}(m) = \frac{m^2 \Gamma(m)}{(m^2 - m_0^2)^2 +m^2\Gamma(m)^2}
\end{equation}
Where $m_0$ is the pole mass and $\Gamma(m)$ is the mass-dependent width. The Manley et al. \cite{PhysRevD.45.4002} approach is used to describe the mass-dependent widths as 
\begin{equation}
\Gamma_{R\to a+ b}(m) = \Gamma^0_{R\to a+ b} \frac{\rho_{ab}(m)}{\rho_{ab}(m_0)}
\end{equation}
where 
\begin{align}
\rho_{ab}(m) = \int dm_a dm_b\mathcal{A}_a(m_a)\mathcal{A}_a(m_b) \frac{\lvert  \vec{p}_f \rvert}{m} \nonumber \\
\cross B_L(\lvert \vec{p}_f \rvert R)\mathcal{F}_{ab}^2(m).
\end{align}
$B_L$ is the Blatt-Weisskopf function, $L$ is the orbital angular momentum of the decay channel, $R$ is the interaction length and $\mathcal{F}_{ab}$ is a form factor for unstable decay products.
The on-shell parameters, branching ratio $\Gamma^0_{R\to a+ b}/\Gamma^0_{\text{tot}}$ , mass $m_0$ and width at the pole is taken from the Particle Data Group (PDG) \cite{PDG}.

\section{Emulation}
Transport codes require significant CPU resources, especially for cross-sections that demand extensive statistics, such as those involving strangeness. Therefore, emulation of transport simulation is preferred. Emulation, in this context, means creating a simplified model that approximates the behavior of the full simulation. A straightforward emulation of strangeness production via resonances involves multiplying the resonance cross-section by the respective branching ratio. By summing all possible resonances, one obtains the portion of the strangeness cross-section due to resonances.

Since SMASH uses mass-dependent partial widths, the branching ratio must be integrated over the possible mass values along with the spectral function. Strings are not affected by resonance parameters; thus, the string cross-section can be simulated once and then added to the emulation. All of the above considerations lead to the following expression:

\begin{align}\label{eq:emul}
\sigma_{d+i} &\approx \sigma_{\text{string} \to d+i} + \sum_{N^*} \sigma_{N^*} \left \langle \frac{\Gamma_{N^* \to d}}{\Gamma_{N^*}} \right\rangle\\
\left \langle \frac{\Gamma_{N^* \to d}}{\Gamma_{N^*}} \right\rangle &=\int_{m_d}^{\sqrt{s} - m_i} \dd m  A(m) \frac{\Gamma_{N^* \to d}(m)}{\Gamma_{N^*}(m)}
\end{align}

where $d$ and $i$ stand for decay products and spectator, respectively. It is important to note if the spectral function is not normalized to the integration interval then the integral in eq. \ref{eq:emul} needs to be divided by 
\begin{equation}
\mathcal{N} = \int_{m_d}^{\sqrt{s} - m_i} \dd m  A(m)
\end{equation}
which ensures that the correct mean for the branching ratio is calculated. For pion-proton collisions that form a resonance through a two-to-one reaction, the mass is not sampled for the mass-dependent branching ratio because the mass is set to $\sqrt{s}$ due to kinematics. 
\section{Data Selection \& Parameter Bounds}
To exclude secondary collisions and other interfering strangeness production interactions, exclusive proton-proton and pion-proton cross-sections were used. The data were collected from the same sources as in \cite{vinzent}, which include \cite{Balewski1998, Sewerin1999, Kowina2004, AbdElSamad2010, AbdelBary2010, Bilger1998, AbdelSamad2006, Baldini1988, Valdau2010, Valdau2007, Ko1983, Wolke1998}.

In principle, any exclusive cross-section can be used, provided that only resonances contribute. However, using multiple cross-sections increases computational complexity, as it requires tuning additional hyperparameters due to the varying weights of each cross-section in the scoring function. Furthermore, as the reward function grows more complex, the algorithm is expected to be less efficient in finding optimal solutions.

To obtain reliable constraints, we selected cross-sections that have similar resonance contributions. For example, the $\Lambda + K^0$ cross-section has resonance contributions similar to the $\Lambda + p + K^+$ cross-section. The cross-sections used in the optimization are the same as those in section \ref{stability}.

For the resonance parameter bounds, we used the limits reported by the PDG whenever available. However, because the default SMASH parameter set includes resonances and decay modes unconfirmed by the PDG, we assigned bounds differently in such cases. Specifically, for the widths and masses of these resonances, we varied the default SMASH values rather than using fixed PDG limits. For unconfirmed decay modes, which are generally small (on the order of $1\%$), we set the default SMASH values as the upper bound and 0 as the lower bound. These choices resulted in the parameter bounds presented in Tables \ref{tab:BR-ref} and \ref{tab:part-ref}
\begin{table}[htbp]
\centering
\begin{tabular}{lccccr}
\toprule
Resonance & Mode & Upper & Lower & SMASH-3.1 \\
\midrule
N(1650) & $\Lambda$ K & 15.0 & 5.0 & 4.0 \\
N(1710) & $\Lambda$ K & 25.0 & 5.0 & 13.0 \\
N(1720) & $\Lambda$ K & 19.0 & 4.0 & 5.0 \\
N(1875) & $\Lambda$ K & 2.0 & 1.0 & 4.0 \\
N(1880) & $\Lambda$ K & 3.0 & 1.0 & 2.0 \\
N(1895) & $\Lambda$ K & 23.0 & 3.0 & 18.0 \\
N(1900) & $\Lambda$ K & 20.0 & 2.0 & 2.0 \\
N(1990) & $\Lambda$ K & 6.1 & 5.9 & 3.0 \\
N(2060) & $\Lambda$ K & 20.0 & 10.0 & 1.0 \\
N(2080) & $\Lambda$ K & 2.0 & 1.0 & 0.5 \\
N(2100) & $\Lambda$ K & 1.0 & 0.0 & 1.0 \\
N(2250) & $\Lambda$ K & 3.0 & 1.0 & 0.2 \\
N(1710) & $\Sigma$ K & 1.0 & 0.0 & 1.0 \\
N(1875) & $\Sigma$ K & 1.1 & 0.3 & 4.0 \\
N(1880) & $\Sigma$ K & 24.0 & 10.0 & 10.0 \\
N(1895) & $\Sigma$ K & 20.0 & 6.0 & 11.0 \\
N(1900) & $\Sigma$ K & 7.0 & 3.0 & 3.0 \\
N(1990) & $\Sigma$ K & 1.0 & 0.0 & 3.0 \\
N(2060) & $\Sigma$ K & 5.0 & 1.0 & 4.0 \\
$\Delta$(1900) & $\Sigma$ K & 1.0 & 0.0 & 1.0 \\
$\Delta$(1910) & $\Sigma$ K & 14.0 & 4.0 & 4.0 \\
$\Delta$(1920) & $\Sigma$ K & 6.0 & 2.0 & 3.0 \\
$\Delta$(1930) & $\Sigma$ K & 1.0 & 0.0 & 4.0 \\
$\Delta$(1950) & $\Sigma$ K & 0.5 & 0.3 & 0.5 \\
N(2080) & N $\phi$ & 1.0 & 0.0 & 1.0 \\
N(2100) & N $\phi$ & 1.0 & 0.0 & 1.0 \\
N(2120) & N $\phi$ & 1.0 & 0.0 & 1.0 \\
N(2190) & N $\phi$ & 1.0 & 0.0 & 1.0 \\
N(2220) & N $\phi$ & 1.0 & 0.0 & 1.0 \\
N(2250) & N $\phi$ & 1.0 & 0.0 & 1.0 \\
N(1880) & N $a_0$(980) & 5.0 & 1.0 & 2.0 \\
$\Delta$(1920) & N $a_0$(980) & 1.0 & 0.0 & 1.0 \\
N(2080) & N $f_0$(980) & 1.0 & 0.0 & 0.1 \\
N(2190) & N $f_0$(980) & 1.0 & 0.0 & 0.1 \\
N(2220) & N $f_0$(980) & 1.0 & 0.0 & 0.1 \\
N(2250) & N $f_0$(980) & 1.0 & 0.0 & 0.1 \\
\bottomrule
\end{tabular}
\caption{Tabulation of branching ratio bounds and current default values for SMASH-3.1}
\label{tab:BR-ref}
\end{table}

\begin{table}[htbp]
\centering
\begin{tabular}{lcccccr}
\toprule
\thead{Resonance} & \thead{Mass[GeV]\\SMASH-3.1} & \thead{Width[GeV]\\SMASH-3.1} & \thead{Mass[GeV]\\Bounds} & \thead{Width[GeV]\\Bounds} \\
\midrule
N(1440) & 1.440 & 0.350 & 1.36-1.38 & 0.18-0.205 \\
N(1520) & 1.515 & 0.110 & 1.505-1.515 & 0.105-0.12 \\
N(1535) & 1.530 & 0.150 & 1.5-1.52 & 0.08-0.13 \\
N(1650) & 1.650 & 0.125 & 1.65-1.68 & 0.1-0.17 \\
N(1675) & 1.675 & 0.145 & 1.65-1.66 & 0.12-0.15 \\
N(1680) & 1.685 & 0.120 & 1.66-1.68 & 0.11-0.135 \\
N(1700) & 1.720 & 0.200 & 1.65-1.75 & 0.1-0.3 \\
N(1710) & 1.710 & 0.140 & 1.65-1.75 & 0.08-0.16 \\
N(1720) & 1.720 & 0.250 & 1.66-1.71 & 0.15-0.3 \\
N(1875) & 1.875 & 0.250 & 1.85-1.95 & 0.1-0.22 \\
N(1880) & 1.880 & 0.400 & 1.82-1.9 & 0.18-0.28 \\
N(1895) & 1.895 & 0.120 & 1.89-1.93 & 0.08-0.14 \\
N(1900) & 1.900 & 0.200 & 1.9-1.94 & 0.09-0.16 \\
N(1990) & 1.990 & 0.500 & 1.9-2.1 & 0.2-0.4 \\
N(2060) & 2.100 & 0.400 & 2.02-2.13 & 0.35-0.43 \\
N(2080) & 2.000 & 0.350 & 1.85-1.92 & 0.12-0.25 \\
N(2100) & 2.100 & 0.260 & 2.05-2.15 & 0.24-0.34 \\
N(2120) & 2.120 & 0.300 & 2.05-2.15 & 0.2-0.36 \\
N(2190) & 2.180 & 0.400 & 1.95-2.15 & 0.3-0.5 \\
N(2220) & 2.220 & 0.400 & 2.13-2.2 & 0.36-0.48 \\
N(2250) & 2.250 & 0.470 & 2.1-2.2 & 0.35-0.5 \\
$\Delta$(1620) & 1.610 & 0.130 & 1.59-1.61 & 0.08-0.14 \\
$\Delta$(1700) & 1.710 & 0.300 & 1.64-1.69 & 0.2-0.3 \\
$\Delta$(1900) & 1.860 & 0.250 & 1.83-1.9 & 0.18-0.3 \\
$\Delta$(1905) & 1.880 & 0.330 & 1.75-1.8 & 0.26-0.34 \\
$\Delta$(1910) & 1.900 & 0.300 & 1.8-1.9 & 0.2-0.5 \\
$\Delta$(1920) & 1.920 & 0.300 & 1.85-1.95 & 0.2-0.4 \\
$\Delta$(1930) & 1.950 & 0.300 & 1.82-1.88 & 0.3-0.45 \\
$\Delta$(1950) & 1.930 & 0.280 & 1.87-1.89 & 0.22-0.26 \\
\bottomrule
\end{tabular}
\caption{Tabulation of the default resonance masses and widths in SMASH-3.1 and bounds used.}
\label{tab:part-ref}
\end{table}

\section{Parameter Stability}\label{stability}

There are significant uncertainties regarding resonance properties. For instance, according to the PDG, the N(1900) resonance has bounds on its decay mode $\Lambda+K$ ranging from $2\%$ to $20\%$, illustrating the extent of these uncertainties. Such large uncertainties in resonance properties can be expected to manifest in observables sensitive to resonances.

To illustrate this, equation \eqref{eq:emul} was used to investigate the parameter stability of exclusive strange cross-sections by varying resonance parameters within the bounds provided by PDG. 

\begin{figure}
\centering
\includegraphics[width=8.6cm]{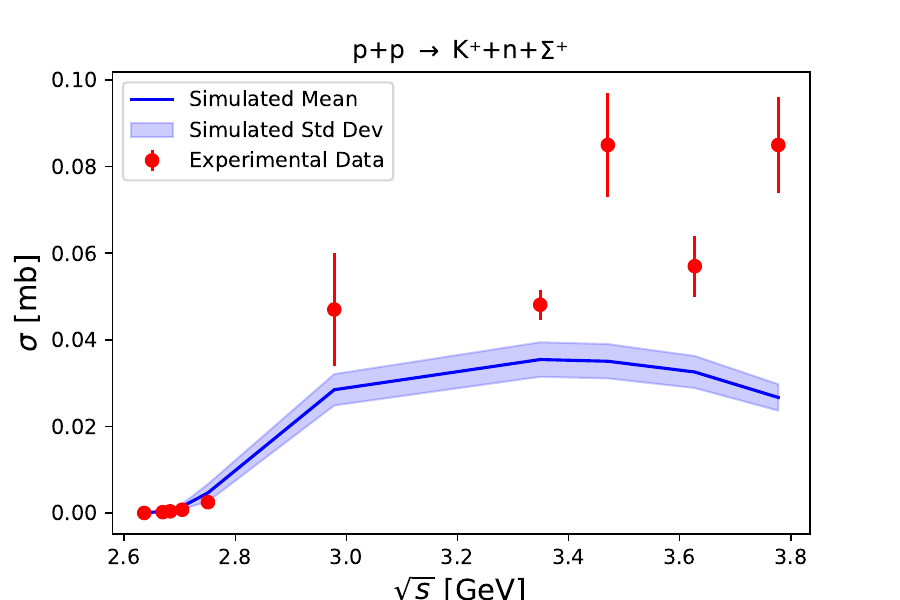}
\caption{Mean exclusive proton-proton cross-section for $K^++n+\Sigma^+$ production as a function of collision energy $\sqrt{s}$, with shaded area representing the standard deviation.}
\label{fig:s-kns}
\end{figure}

\begin{figure}
\centering
\includegraphics[width=8.6cm]{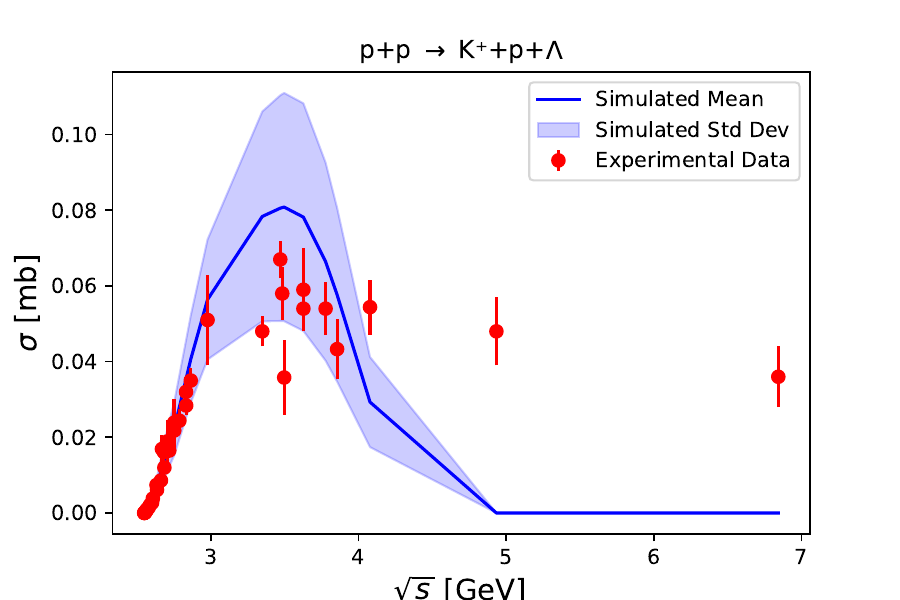}
\caption{Mean exclusive proton-proton cross-section for $K^++p+\Lambda$ production as a function of collision energy $\sqrt{s}$, with shaded area representing the standard deviation.}
\label{fig:s-kpl}
\end{figure}

\begin{figure}
\centering
\includegraphics[width=8.6cm]{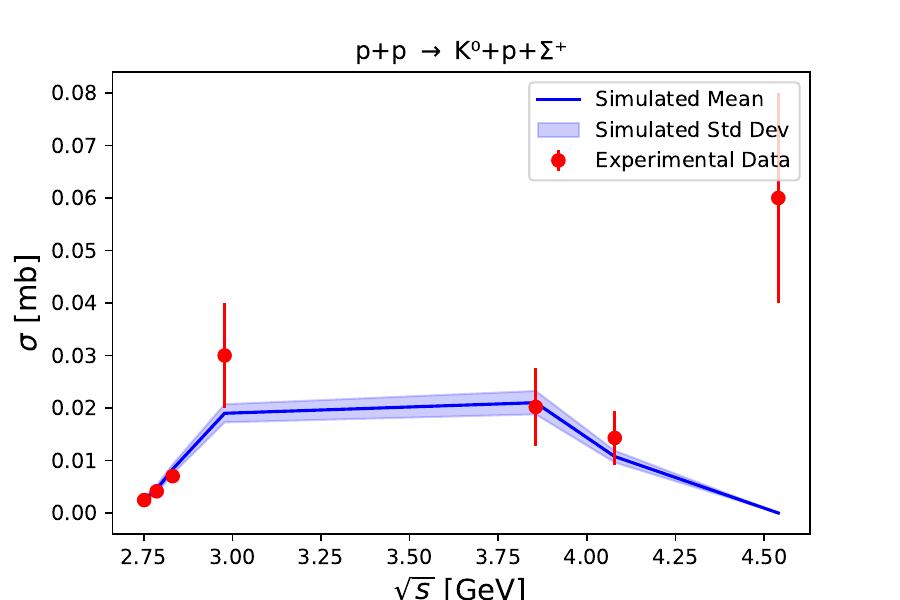}
\caption{Mean exclusive proton-proton cross-section for $K^0+p+\Sigma^+$ production as a function of collision energy $\sqrt{s}$, with shaded area representing the standard deviation.}
\label{fig:s-k0S+}
\end{figure}

\begin{figure}
\centering
\includegraphics[width=8.6cm]{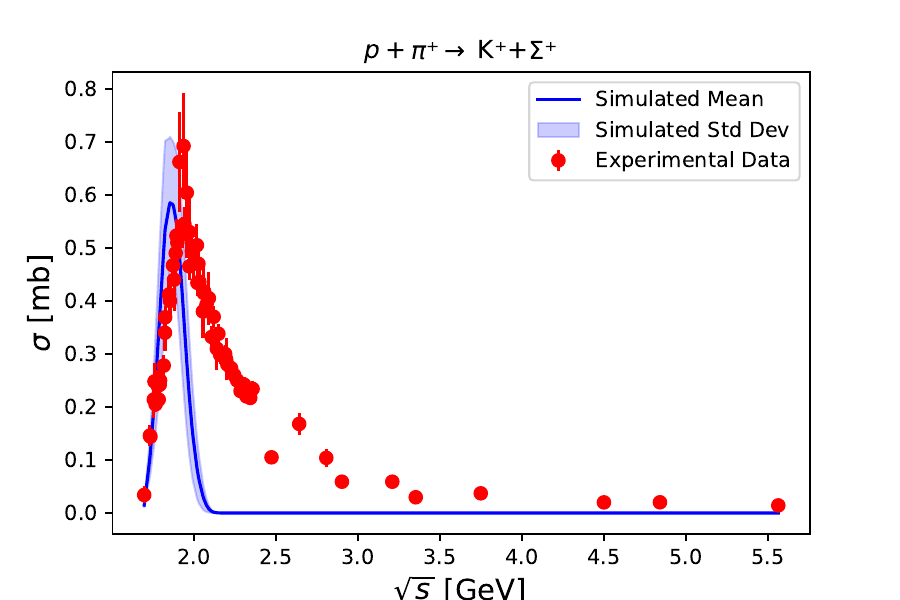}
\caption{Mean exclusive  $\pi^+$-proton cross-section for $K^++\Sigma^+$ production as a function of collision energy $\sqrt{s}$, with shaded area representing the standard deviation.}
\label{fig:s-k+S+}
\end{figure}

\begin{figure}
\centering
\includegraphics[width=8.6cm]{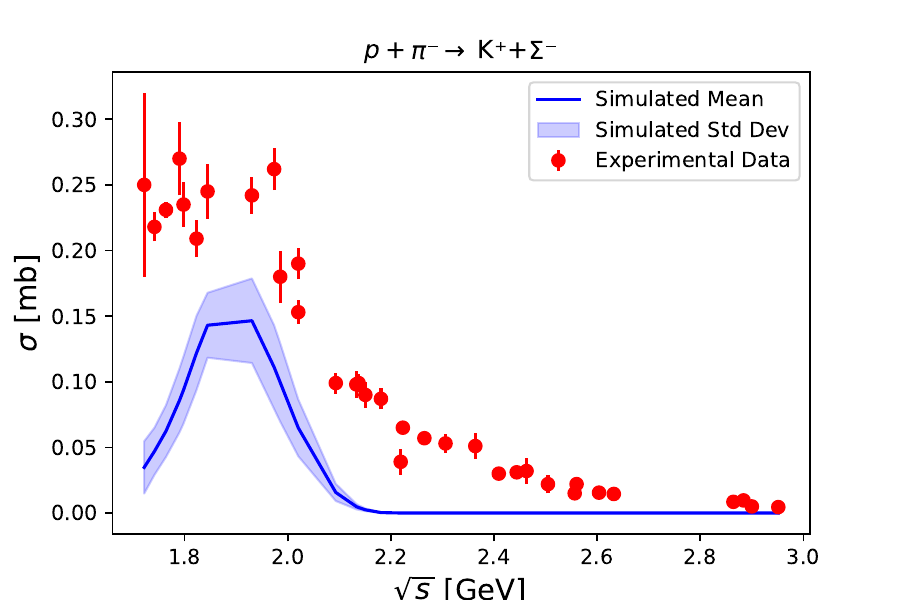}
\caption{Mean exclusive $\pi^-$-proton cross-section for $K^++\Sigma^-$ production as a function of collision energy $\sqrt{s}$, with shaded area representing the standard deviation.}
\label{fig:s-k+S-}
\end{figure}

\begin{figure}
\centering
\includegraphics[width=8.6cm]{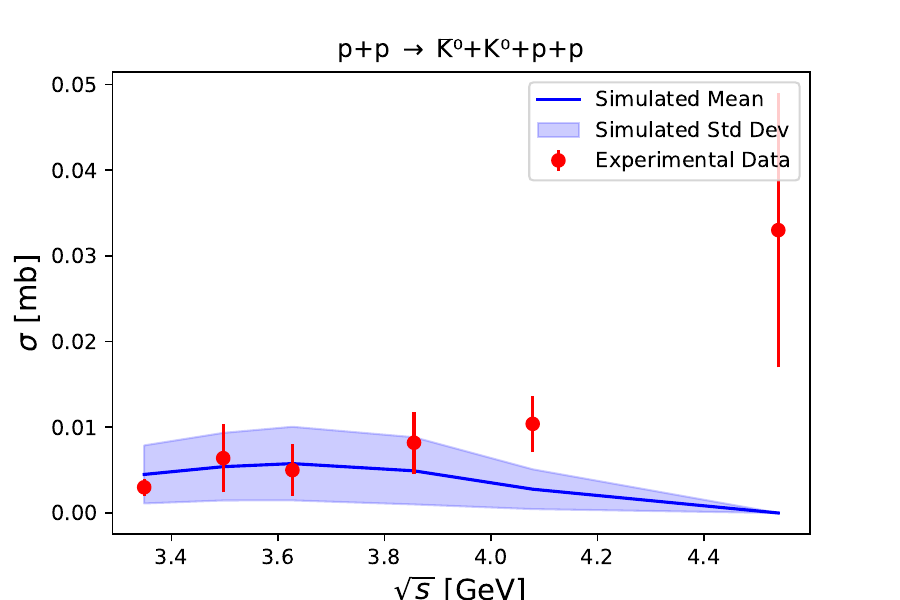}
\caption{Mean exclusive proton-proton cross-section for $p+p+K^0+\bar{K}^0$ production as a function of collision energy $\sqrt{s}$, with shaded area representing the standard deviation.}
\label{fig:s-k0k0pp}
\end{figure}

\begin{figure}
\centering
\includegraphics[width=8.6cm]{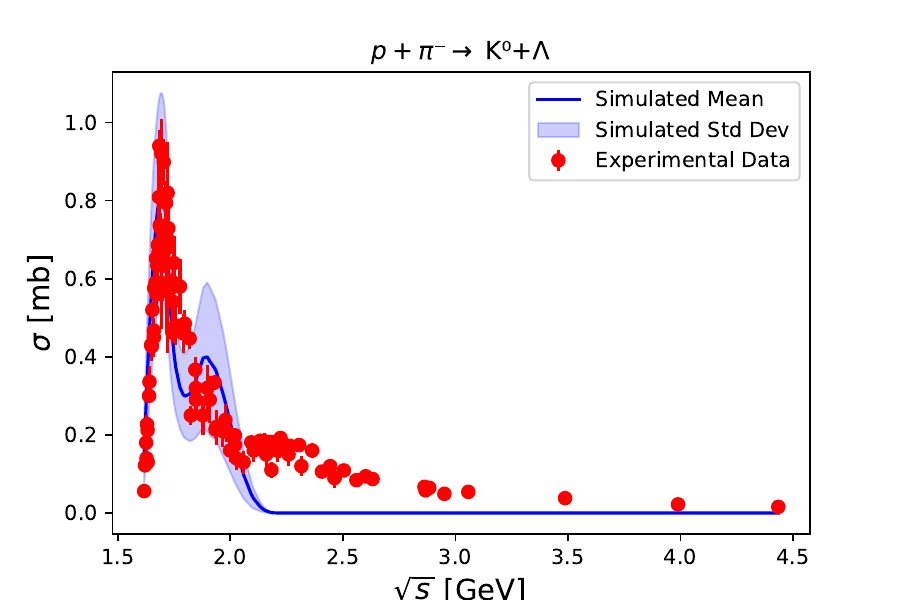}
\caption{Mean exclusive $\pi^-$-proton cross-section for $K^0+\Lambda$ production as a function of collision energy $\sqrt{s}$, with shaded area representing the standard deviation.}
\label{fig:s-k0L}
\end{figure}

Figures \ref{fig:s-kns}, \ref{fig:s-kpl}, \ref{fig:s-k0S+}, \ref{fig:s-k+S+}, \ref{fig:s-k+S-}, \ref{fig:s-k0k0pp}, and \ref{fig:s-k0L} show the mean respective cross-sections and standard deviations generated by running the emulator without strings for randomized parameter sets. These figures demonstrate that some cross-sections, such as \ref{fig:s-k0S+}, are more stable compared to others like \ref{fig:s-kpl}, which exhibit larger standard deviations relative to their means.
More importantly, the means of some cross-sections are closer to the experimental data than others. Comparing Figure \ref{fig:s-k+S-} to Figure \ref{fig:s-k0S+} highlights the difference in closeness to the experimental data. The mean cross-section in Figure \ref{fig:s-k+S-}, particularly in the lower energy region, deviates by several standard deviations from the experimental data points.

\section{Genetic Algorithm}

Genetic Algorithms (GAs) are a class of optimization algorithms inspired by the principles of natural selection and genetics. They excel in solving complex problems where traditional optimization techniques struggle, such as noisy black-box functions lacking derivative information. In such scenarios, stochastic algorithms may find it challenging to navigate multiple optima effectively. Furthermore, GAs are embarrassingly parallelizable leading allowing for efficient use of CPU resources.

Genetic Algorithms (GAs) belong to the broader family of evolutionary algorithms, which mimic natural evolution to find optimal or near-optimal solutions. These population-based optimization techniques utilize Darwinian principles to iteratively search for optimal solutions.

\subsection{Algorithm}

A GA operates through several key steps, each crucial to its iterative optimization process:

\begin{itemize}
    \item \textbf{Initialization}: Start by creating an initial population $P$ of random solutions. Each solution represents a potential answer to the problem, often encoded as a binary string or an array of parameters.
    
    \item \textbf{Evaluation and Fitness}: Assess the fitness of each individual in population $P$ by applying a fitness function. This function quantifies how well each solution performs in solving the problem. Higher fitness scores indicate better solutions.
    
    \item \textbf{Evolution Loop}: Iterate until a termination criterion is met, such as reaching a desired fitness level or a maximum number of generations:
    
    \begin{itemize}
        \item \textbf{Selection}: Choose individuals from population $P$ to act as parents for the next generation's offspring. Selection methods typically favor individuals with higher fitness scores, using techniques like tournament selection or roulette wheel selection.
        
        \item \textbf{Crossover}: Perform crossover, a genetic operator where pairs of parent solutions exchange genetic information to create new offspring. This process mimics biological recombination and aims to combine beneficial traits from different solutions. In this paper, a random uniform crossover is used, where each bit or parameter of the offspring is randomly inherited from either parent.
        
        \item \textbf{Mutation}: Introduce diversity by applying mutation to offspring solutions. Mutation randomly alters a small portion of offspring genes, helping to explore new regions of the solution space that might lead to better solutions. The mutation rate governs the probability of mutation occurrence.
        
        \item \textbf{Evaluation of Offspring}: Assess the fitness of newly generated offspring using the same fitness function as before.
        
        \item \textbf{Survivor Selection}: Combine the current population $P$ with the offspring. Then, select individuals to form the next generation, typically favoring higher fitness individuals to ensure that the population continues to improve over generations.
    \end{itemize}
\end{itemize}

This paper employs a straightforward approach to the genetic algorithm, following the outlined steps. Specifically, crossover involves a random mixing of two solutions, while mutation introduces random alterations to the offspring solutions. The chosen mutation rate was 10\%, meaning each parameter has a 10\% chance of being randomized. Furthermore, the selection process eliminates the bottom 50\% of solutions in each generation. However, many advanced techniques exist to enhance specific algorithmic steps, such as probabilistic selection methods and more sophisticated crossover strategies tailored to the problem domain.

\subsection{Fitness Function}
The purpose of the fitness function is to quantify the quality of a solution. In our case, the quality of a solution depends on how well the simulation reproduces experimental data. Hence, the fitness function should measure how closely the simulation matches experimental results. A simple choice is to define it in terms of the relative difference between the simulation and experimental data:
\begin{equation}
D(\sqrt{s},\text{solution}) = \frac{\left|\sigma_{\text{solution}}(\sqrt{s}) - \sigma_{\text{exp}}(\sqrt{s})\right|}{\sigma_{\text{exp}}(\sqrt{s})}.
\end{equation}
To calculate the fitness, we sum over all data points. When incorporating multiple cross-sections, we must determine how to weigh them, as each cross-section may contain a different number of data points. Should cross-sections with more data points carry more weight? This naturally occurs if one cross-section has more data points. However, by using the mean for each cross-section instead of just summing the data, we ensure that each cross-section contributes equally to the scoring. Additionally, a weight based on the error of each data point is applied:

\begin{equation}
W(\sqrt{s},\text{solution})=  \frac{\sigma_{\text{exp}}(\sqrt{s}) - \sigma_{\text{exp,error}}(\sqrt{s})}{\sigma_{\text{exp}}(\sqrt{s})}.
\end{equation}
This results in a fitness function defined as

\begin{equation}\label{eq:fit}
F(\text{solution}) = \frac{1}{N_\text{points}}\sum_{\sqrt{s}} W(\sqrt{s},\text{solution})D(\sqrt{s},\text{solution}) 
\end{equation}

where $N_\text{points}$ is the number of data points for each cross-section. Equation \ref{eq:fit} can then be used to calculate the fitness for each cross-section. However, following the logic from section \ref{stability} about parameter stability, there will be a bias towards parameters that improve the cross-sections which are, on average, close to experimental data. To allow parameter sets that improve the cross-sections which are, on average, further from the data, weights on each cross-section were used. These weights were easily determined by running the algorithm with a small population (for time efficiency) and tuning by hand. Then, the "score" of a solution can be defined as the reciprocal of its total fitness, allowing for a more natural comparison between solutions.

\section{Results \& Discussion}

\begin{figure}
\centering
\includegraphics[width=8.6cm]{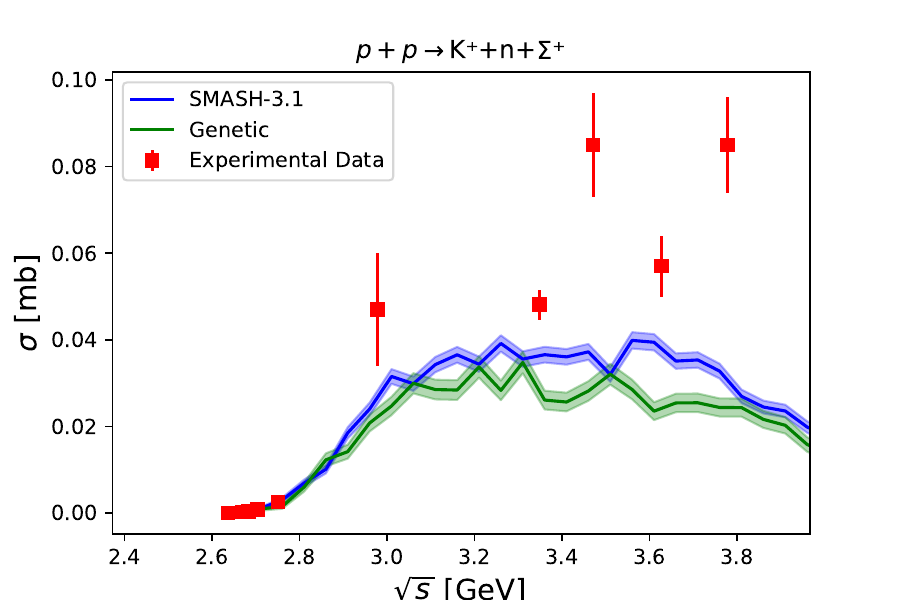}
\caption{Exclusive proton-proton cross-section for $K^++n+\Sigma^+$ production as a function of collision energy $\sqrt{s}$.}
\label{fig:kns}
\end{figure}

\begin{figure}
\centering
\includegraphics[width=8.6cm]{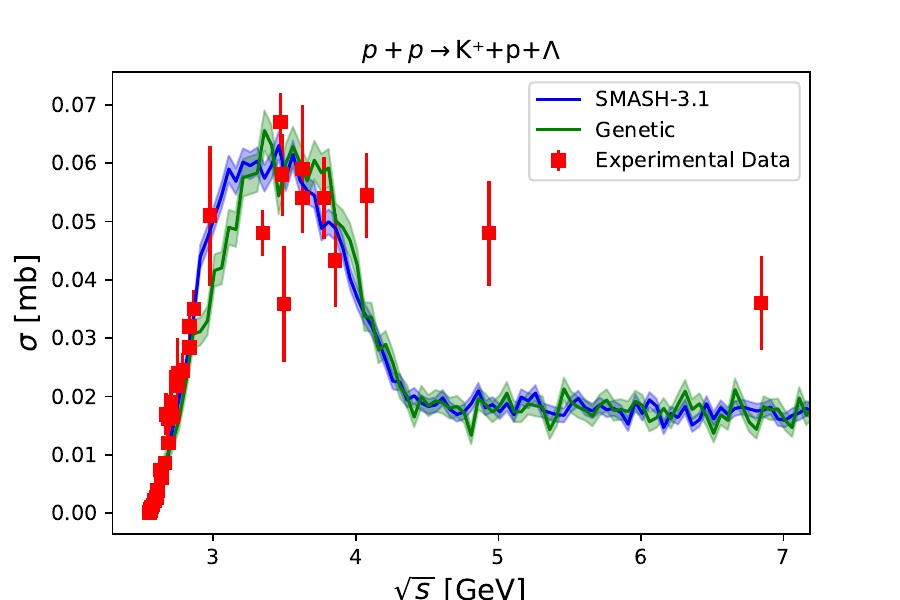}
\caption{Exclusive proton-proton cross-section for $K^++p+\Lambda$ production as a function of collision energy $\sqrt{s}$.}
\label{fig:kpl}
\end{figure}

\begin{figure}
\centering
\includegraphics[width=8.6cm]{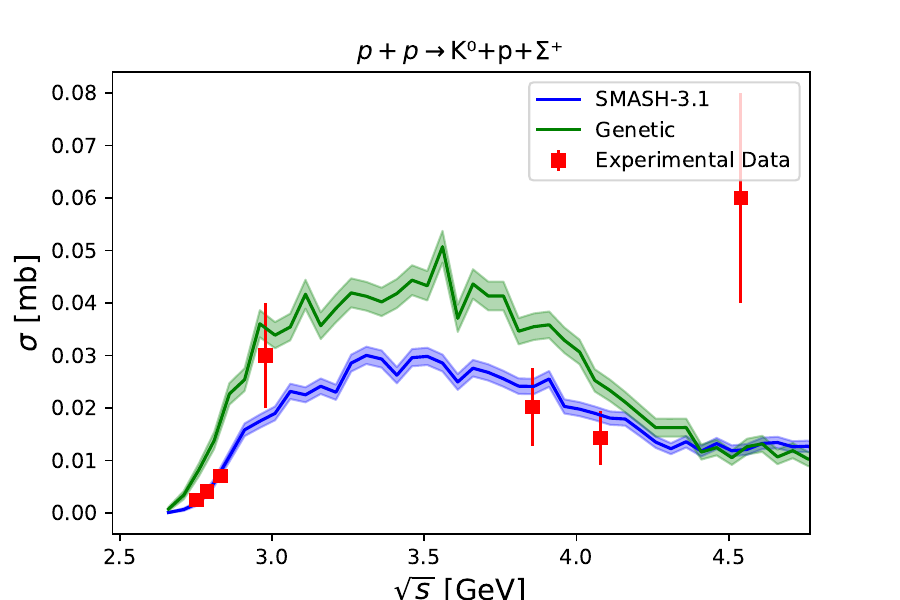}
\caption{Exclusive proton-proton cross-section for $K^0+p+\Sigma^+$ production as a function of collision energy $\sqrt{s}$.}
\label{fig:k0pS+}
\end{figure}

\begin{figure}
\centering
\includegraphics[width=8.6cm]{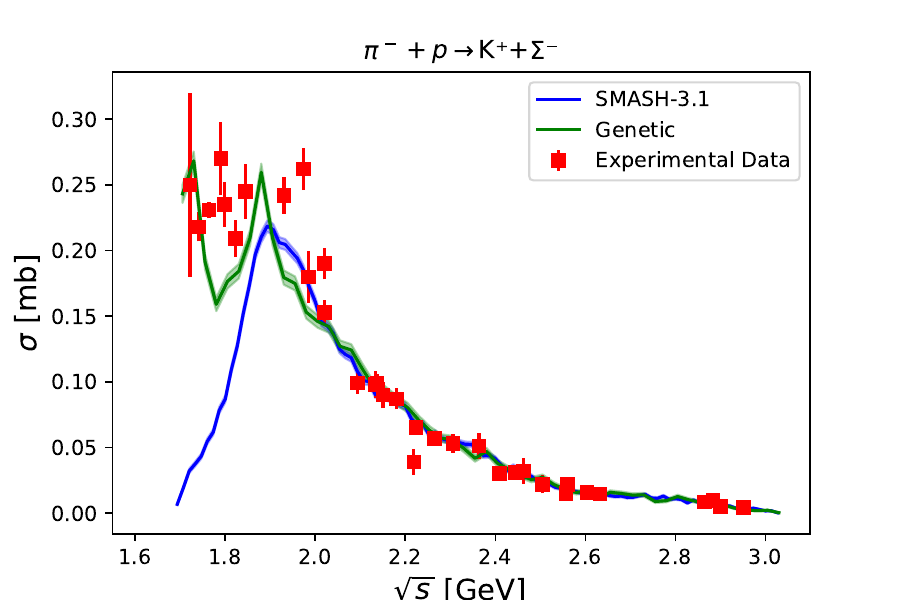}
\caption{Exclusive $\pi^-$-proton cross-section for $K^++\Sigma^-$ production as a function of collision energy $\sqrt{s}$.}
\label{fig:k+S-}
\end{figure}

Tables \ref{tab:BR} and \ref{tab:part} compare the results from the genetic algorithm optimization with the default parameters of the SMASH transport model. While the default masses and widths are similar to the fitted values, the branching ratios show significant changes.
For the proton-proton cross-sections, the results shown in Figs. \ref{fig:kns} and \ref{fig:kpl} are similar to the default SMASH parameters. In contrast, Fig. \ref{fig:k0pS+} shows a worse fit to the experimental data compared to the default parameters. However, for the pion-proton cross-sections, the algorithm identifies an alternative solution that is more faithful to experimental data, as seen in Fig. \ref{fig:k+S-}. Hence, there is some tension between the cross-sections in Figs. \ref{fig:k0pS+} and \ref{fig:k+S-}. 
\\

The pion-proton cross-sections exhibit dips at certain collision energies that are not observed in the experimental data. Since the genetic algorithm did not find a solution that resolves these dips, it suggests that they may be caused by missing resonances or overly restrictive mass bounds in the optimization. Pion-proton cross-sections are more sensitive to resonance masses compared to proton-proton collisions because, in pion-proton interactions, the resonance mass is determined by kinematics rather than being directly sampled. This leads to narrower peaks that are more sensitive to parameter variations. For instance, the peak observed in Fig. \ref{fig:k+S-} is attributed to the decay of the $N(1710)$ resonance.
\\

\begin{figure}
\centering
\includegraphics[width=8.6cm]{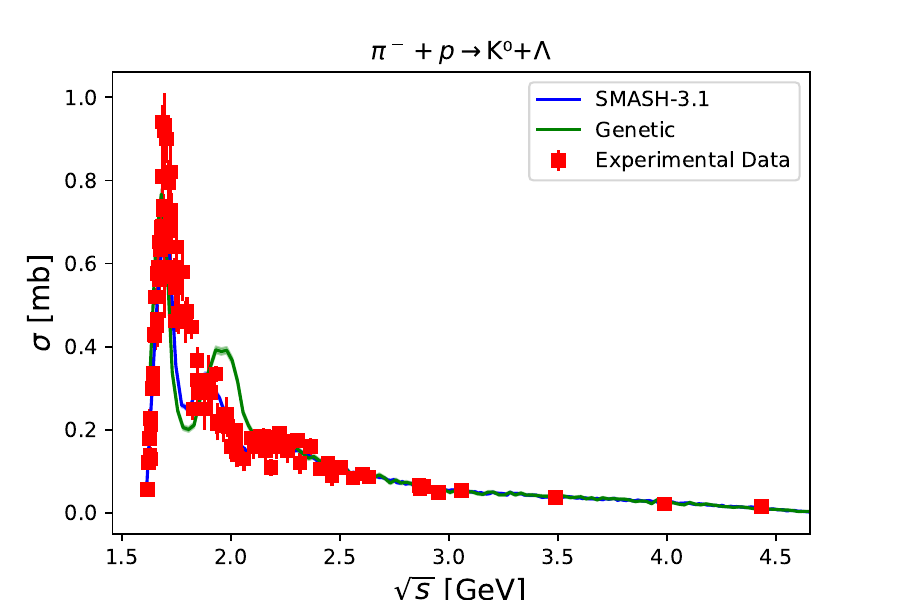}
\caption{Exclusive $\pi^-$-proton cross-section for $K^0+\Lambda$ production as a function of collision energy $\sqrt{s}$.}
\label{fig:k0S+}
\end{figure}

A notable feature observed in Fig. \ref{fig:k0S+} is a bump around $\sqrt{s} \approx \SI{2}{GeV}$, which can be attributed to the N(1990) and N(2060) resonances. The positions of these peaks are determined by the resonance pole masses, while their heights depend on the branching ratios. The default SMASH parameters, which use older PDG bounds for the branching ratios, result in much lower values, and consequently, no bump is observed in the cross-section. The genetic algorithm, which incorporates newer PDG values for the branching ratios, yields higher values for the N(1990) and N(2060) resonances, leading to the observed bump in the cross-section. Updating the SMASH to incorporate the newer branching ratio bounds would likely result in the default parameter set producing the same bump in the cross-section, as shown in Fig. \ref{fig:k0S+}.

\begin{table}[htbp]
\centering
\begin{tabular}{lccccr}
\toprule
\thead{Resonance} & \thead{Mode} & \thead{Br[\%]\\(Default)} & \thead{Br[\%]\\(Genetic)} \\
\midrule
N(1650) & $\Lambda$ K & 4.000 & 8.289 \\
N(1710) & $\Lambda$ K & 13.000 & 5.107 \\
N(1710) & $\Sigma$ K & 1.000 & 0.995 \\
N(1720) & $\Lambda$ K & 5.000 & 6.165 \\
N(1875) & $\Sigma$ K & 4.124 & 0.961 \\
N(1875) & $\Lambda$ K & 4.124 & 1.042 \\
N(1880) & $\Lambda$ K & 2.000 & 1.021 \\
N(1880) & $\Sigma$ K & 10.000 & 23.813 \\
N(1880) & N $a_0$(980) & 2.000 & 3.252 \\
N(1895) & $\Lambda$ K & 18.000 & 3.738 \\
N(1895) & $\Sigma$ K & 11.000 & 19.925 \\
N(1900) & $\Lambda$ K & 2.000 & 2.004 \\
N(1900) & $\Sigma$ K & 3.000 & 5.690 \\
N(1990) & $\Lambda$ K & 3.000 & 5.912 \\
N(1990) & $\Sigma$ K & 3.000 & 0.776 \\
N(2060) & $\Lambda$ K & 1.000 & 10.025 \\
N(2060) & $\Sigma$ K & 4.000 & 4.972 \\
N(2080) & $\Lambda$ K & 0.558 & 1.143 \\
N(2080) & N $\phi$ & 1.116 & 0.600 \\
N(2080) & N $f_0$(980) & 0.112 & 0.760 \\
N(2100) & $\Lambda$ K & 0.990 & 0.477 \\
N(2100) & N $\phi$ & 0.990 & 0.364 \\
N(2120) & N $\phi$ & 0.988 & 0.341 \\
N(2190) & N $\phi$ & 1.000 & 0.992 \\
N(2190) & N $f_0$(980) & 0.100 & 0.335 \\
N(2220) & N $\phi$ & 1.122 & 0.303 \\
N(2220) & N $f_0$(980) & 0.112 & 0.547 \\
N(2250) & $\Lambda$ K & 0.226 & 1.080 \\
N(2250) & N $\phi$ & 1.129 & 0.928 \\
N(2250) & N $f_0$(980) & 0.113 & 0.149 \\
$\Delta$(1900) & $\Sigma$ K & 1.000 & 0.589 \\
$\Delta$(1910) & $\Sigma$ K & 4.000 & 5.384 \\
$\Delta$(1920) & $\Sigma$ K & 3.000 & 5.816 \\
$\Delta$(1920) & N $a_0$(980) & 1.000 & 0.033 \\
$\Delta$(1930) & $\Sigma$ K & 4.000 & 0.279 \\
$\Delta$(1950) & $\Sigma$ K & 0.500 & 0.449 \\
\bottomrule
\end{tabular}
\caption{Tabulation of the default branching ratios of SMASH-3.1 and the result of the genetic algorithm.}
\label{tab:BR}
\end{table}

\begin{table}[htbp]
\centering
\begin{tabular}{lcccccr}
\toprule
\thead{Resonance} & \thead{Mass[Gev]\\(Default)} & \thead{Mass[Gev]\\(Genetic)} & \thead{Width[Gev]\\(Default)} & \thead{Width[Gev]\\(Genetic)} \\
\midrule
N(1650) & 1.650 & 1.651 & 0.125 & 0.119 \\
N(1710) & 1.710 & 1.685 & 0.140 & 0.158 \\
N(1720) & 1.720 & 1.680 & 0.250 & 0.152 \\
N(1875) & 1.875 & 1.855 & 0.250 & 0.209 \\
N(1880) & 1.880 & 1.823 & 0.400 & 0.246 \\
N(1895) & 1.895 & 1.890 & 0.120 & 0.130 \\
N(1900) & 1.900 & 1.901 & 0.200 & 0.157 \\
N(1990) & 1.990 & 2.039 & 0.500 & 0.379 \\
N(2060) & 2.100 & 2.128 & 0.400 & 0.429 \\
N(2080) & 2.000 & 1.852 & 0.350 & 0.226 \\
N(2100) & 2.100 & 2.115 & 0.260 & 0.339 \\
N(2120) & 2.120 & 2.118 & 0.300 & 0.354 \\
N(2190) & 2.180 & 2.090 & 0.400 & 0.328 \\
N(2220) & 2.220 & 2.137 & 0.400 & 0.460 \\
N(2250) & 2.250 & 2.177 & 0.470 & 0.393 \\
$\Delta$(1900) & 1.860 & 1.848 & 0.250 & 0.207 \\
$\Delta$(1910) & 1.900 & 1.896 & 0.300 & 0.462 \\
$\Delta$(1920) & 1.920 & 1.945 & 0.300 & 0.226 \\
$\Delta$(1930) & 1.950 & 1.848 & 0.300 & 0.336 \\
$\Delta$(1950) & 1.930 & 1.880 & 0.280 & 0.223 \\
\bottomrule
\end{tabular}
\caption{Tabulation of the default resonance masses and widths in SMASH-3.1 and the result of the genetic algorithm.}
\label{tab:part}
\end{table}


\begin{figure}
\centering
\includegraphics[width=8.6cm]{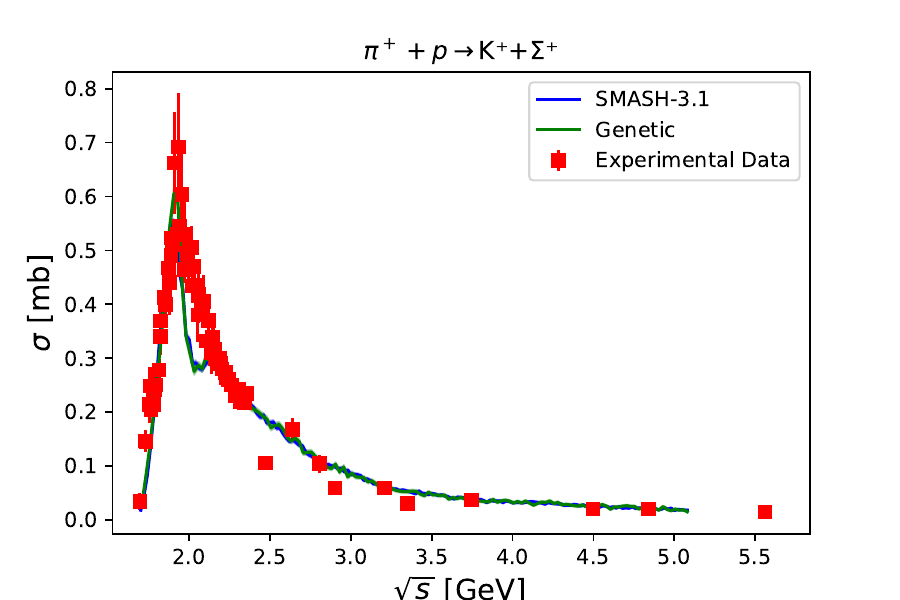}
\caption{Exclusive $\pi^+$-proton cross-section for $K^++\Sigma^+$ production as a function of collision energy $\sqrt{s}$.}
\label{fig:k+S+}
\end{figure}
\begin{figure}
\centering
\includegraphics[width=8.6cm]{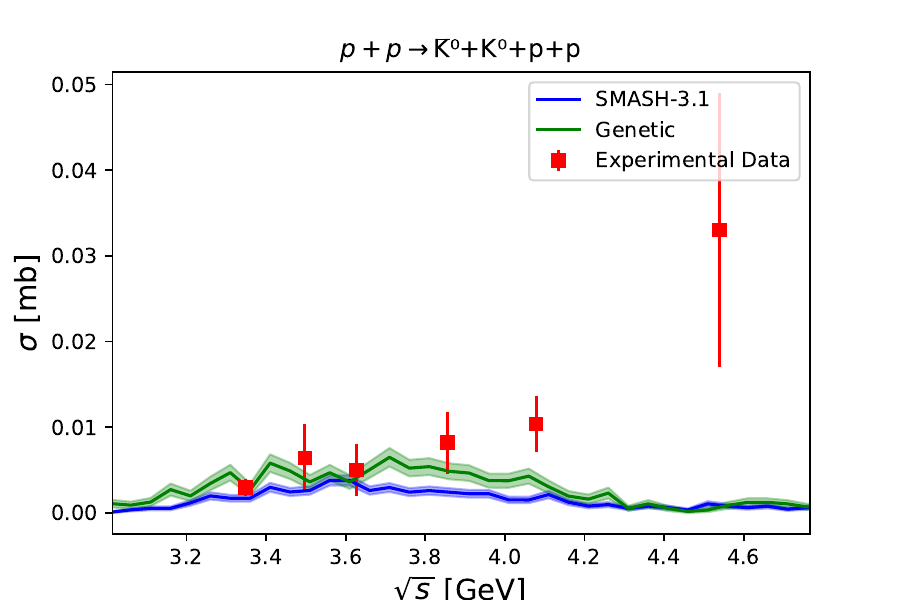}
\caption{Exclusive proton-proton cross-section for $\bar{K}^0 + K^0 +p + p$ production as a function of collision energy $\sqrt{s}$.}
\label{fig:k0k0pp}
\end{figure}

\section{Conclusion \& Outlook}
In this study, we optimized strangeness production from resonances within the SMASH transport model. By applying a genetic algorithm, we systematically adjusted resonance parameters to better align with experimental data, addressing the substantial uncertainties in branching ratios, masses, and widths as reported by the PDG.

Our findings show that, while the default SMASH parameters provide a reasonable baseline, significant improvements can be achieved through optimization. The genetic algorithm effectively navigated the complex parameter space, identifying alternative sets of branching ratios and resonance properties that more accurately reproduce experimental cross-sections, particularly for pion-proton interactions. Therefore, the parameter set from the genetic algorithm was incorporated into the SMASH-3.2 \cite{weil_2025_14922777} as given in tables \ref{tab:BR} and \ref{tab:part}. The properties of two nucleon resonances are kept at their SMASH-3.1 default values, the N(2080) and N(1720) specifically, to ensure agreement with dilepton production that constrains the $\phi$ meson production yields \cite{Steinberg:2018yoj}. 
The sensitivity of pion-proton cross-sections to resonance masses, with sharp peaks resulting from kinematic constraints, underscores the importance of precise parameter tuning in modeling strangeness production. However, the fit was worse for one case, as shown in Fig. \ref{fig:k0pS+}, suggesting some tension between the observables. Therefore, applying weights to the cross-sections or energy intervals contributing to the score could help resolve this, and exploring alternative scoring functions may lead to further improvements.

Despite these challenges, the algorithm presented here is efficient in terms of CPU usage and computation time, providing a strong foundation for future optimization algorithms. Future studies could include additional resonances and use more relaxed bounds for a broader exploration, potentially shedding light on poorly constrained resonance properties.

Overall, the approach and findings demonstrate the potential for significantly improving the accuracy of resonance-based transport models. By fine-tuning model parameters using optimization algorithms, we can achieve a closer alignment with experimental data, leading to more reliable simulations and a better understanding of the underlying physics in heavy-ion collisions.

\section{Acknowledgments}
C. B. Rosenkvist acknowledges fruitful discussions with R. Góes-Hirayama and J. Mohs.
Computational resources have been provided by the Center for Scientific Computing (CSC) at the Goethe-University of Frankfurt.
C. B. Rosenkvist acknowledges financial support from the F\&E program of GSI. \

\bibliography{genetic}

\end{document}